\documentclass[osajnl,twocolumn,showpacs,superscriptaddress,10pt]{revtex4-1} 
\usepackage{amsmath,amssymb,graphicx}

\usepackage{mathtools}
\usepackage{xfrac}

\begin{document}

\title{Coherent, directional supercontinuum via cascaded dispersive wave generation}

\author{Yoshitomo Okawachi}\email{Corresponding author: y.okawachi@columbia.edu}
\affiliation{Department of Applied Physics and Applied Mathematics, Columbia University, New York, NY 10027}

\author{Mengjie Yu}
\affiliation{Department of Applied Physics and Applied Mathematics, Columbia University, New York, NY 10027}
\affiliation{School of Electrical and Computer Engineering, Cornell University, Ithaca, NY 14853}

\author{Jaime Cardenas}
\affiliation{Department of Electrical Engineering, Columbia University, New York, NY 10027}
\affiliation{current address: The Institute of Optics, University of Rochester, Rochester, NY 14627}

\author{Xingchen Ji}
\affiliation{School of Electrical and Computer Engineering, Cornell University, Ithaca, NY 14853}
\affiliation{Department of Electrical Engineering, Columbia University, New York, NY 10027}

\author{Michal Lipson}
\affiliation{Department of Electrical Engineering, Columbia University, New York, NY 10027}

\author{Alexander L. Gaeta}
\affiliation{Department of Applied Physics and Applied Mathematics, Columbia University, New York, NY 10027}

\begin{abstract}We demonstrate a novel approach to producing coherent, directional supercontinuum via cascaded dispersive wave generation. By pumping in the normal group-velocity dispersion regime, pulse compression of the first dispersive wave results in the generation of a second dispersive wave, resulting in an octave-spanning supercontinuum generated primarily to one side of the pump spectrum. We theoretically investigate the dynamics and show that the generated spectrum is highly coherent. We experimentally confirm this dynamical behavior and the coherence properties in silicon nitride waveguides by performing direct detection of the carrier-envelope-offset frequency of our femtosecond pump source using an \emph{f}-2\emph{f} interferometer. Our technique offers a path towards a stabilized, high-power, integrated supercontinuum source with low noise and high coherence, with applications including direct comb spectroscopy.     
\end{abstract}


\maketitle 


\begin{figure}[t]
\centering
\centerline{\includegraphics[width=8.5cm]{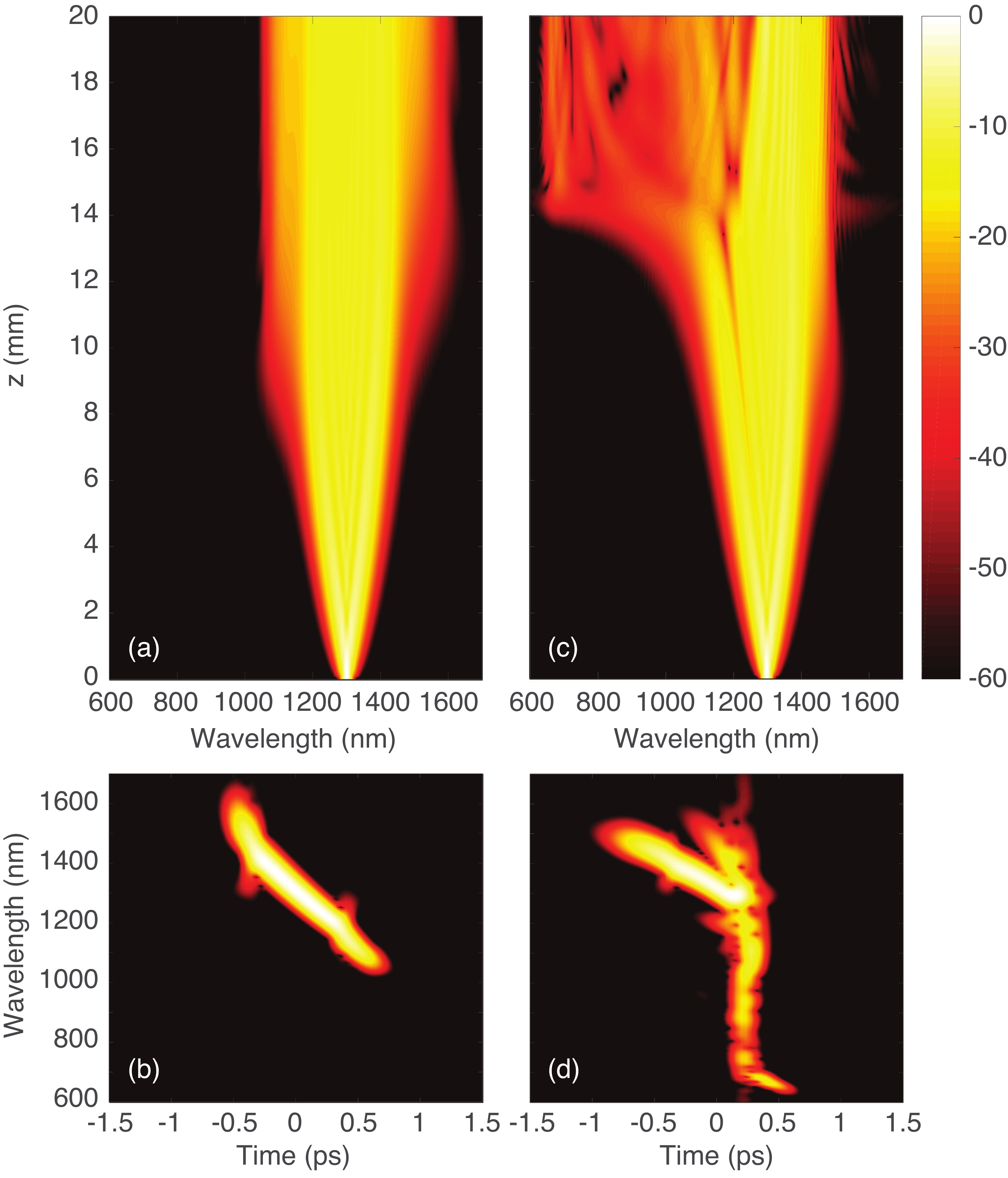}}
\caption{Simulated SCG in a 2-cm Si$_3$N$_4$ waveguide with a cross section of 730$\times$700 nm. The pump wavelength is 1300 nm which is in the normal GVD regime. The left side shows the spectral evolution (a) and the output spectrogram considering only the effects of second-order dispersion. The right side shows the spectral evolution (c) and the output spectrogram (d) considering all-orders of dispersion.}
\label{Fig1}
\end{figure}

Over the past decade, there have been numerous developments of chip-based supercontinuum (SC) sources \cite{Lamont, Phillips, Kuyken, Duchesne, Halir, Epping, Lau, Yu, Leo, Kuyken15, Mayer, Johnson, Klenner, Liu, Oh, Porcel, Hickstein, Herkommer}. For many applications that require a stabilized frequency comb source, the generated spectrum must be phase coherent, allowing for carrier-envelope-offset frequency ($f\textsubscript{CEO}$) detection and self-referencing using an \emph{f}-2\emph{f} interferometer. Such supercontinuum generation (SCG) enables the realization of a range of applications including optical clocks, frequency metrology, pulse compression, and spectroscopy \cite{Dudley}. Silicon-based platforms, have drawn interest as a path towards a complementary metal-oxide-semiconductor (CMOS) process compatible, integrated SC source \cite{Kuyken, Duchesne, Halir, Epping, Lau, Leo, Kuyken15, Mayer, Johnson, Klenner, Liu, Oh, Porcel, Hickstein, Herkommer}. In addition, such platforms allow for high optical confinement due to the high index contrast between the waveguide and the cladding, which yields a large effective nonlinearity and the ability to tailor the dispersion of the waveguide \cite{Turner, Okawachi} and allows for the study of a wide range of nonlinear interactions with moderate pump powers. 

One of the effects that often accompanies SCG is dispersive wave (DW) generation \cite{Efimov,Dudley}. DW generation is a phase-matched process that depends on higher-order dispersion (HOD) effects and provides a means for transferring the energy to wavelengths that are far from the pump, across a zero group-velocity dispersion (GVD) point. This has been utilized for generating broadband SCG \cite{Leo14} and has been used to study more exotic phenomena such as rogue waves \cite{DemircanRogue, DudleyPhoton} and the event horizon \cite{WebbNC,Ciret}. Much of the related work has focused on the interaction between the soliton and a DW, in which the pump is located in the anomalous-GVD regime \cite{Roy10}. Alternatively, several studies have explored pumping in the normal-GVD regime in photonic crystal fibers. However, the red-shifted DW relies on the Raman effect and in some cases a weak probe that is located at the Raman frequency \cite{Qiu, Roy}. Recent theoretical simulations have shown that it is possible to excite DW's pumping in the normal-GVD regime \cite{Webb}. 

In this paper, we investigate theoretically and experimentally SCG via pumping in the normal-GVD regime which allows for a spectrum generated primarily to one side of the input pulse spectrum. Through tailoring the dispersion of a Si$_3$N$_4$ waveguide, we achieve cascaded DW formation across two zero-GVD points, where pulse compression of the first DW results in the generation of a second DW and results in a SC generated primarily to the blue. We demonstrate a 1.2-octave-spanning SCG pumping at 1300 nm that spans 657 -- 1513 nm (30 dB bandwidth). We also investigate the coherence of the generated SC spectrum and show direct detection of the $f\textsubscript{CEO}$ using an \emph{f}-2\emph{f} interferometer. The ability to stabilize the spectrum via self-referencing while pumping in the normal-GVD regime offers advantages since spectral broadening in this regime results in lower noise and higher coherence \cite{Millot}. 

\begin{figure}[tbp]
\centerline{\includegraphics[width=8.5cm]{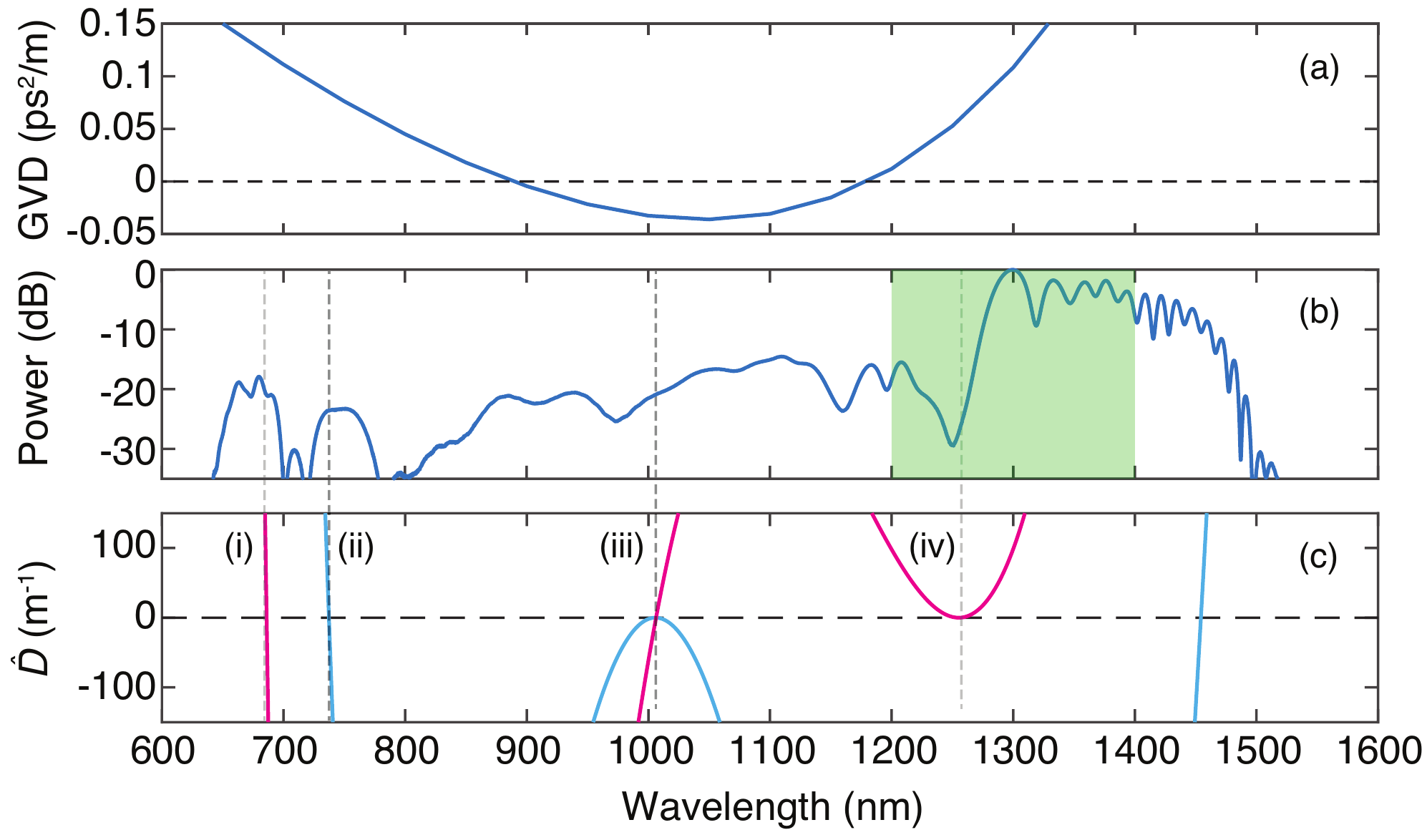}}
\caption{(a) Simulated GVD of 730$\times$700 nm Si$_3$N$_4$ waveguide. (b) Simulated spectrum via pumping at 1300 nm. The spectrum shows two distinct dispersive wave peaks near 685 nm and 740 nm. These peaks arise from two different processes as indicated by the dispersion operators (c) for two different pump wavelengths. The spectral component at (iv) 1255 nm is depleted as it generates dispersive wave components near (i) 685 nm and (iii) 1 $\mu$m [$z =$ 14 mm in Fig.  \ref{Fig1}(c)]. The spectral component near 1 $\mu$m subsequently generates the second dispersive wave near (ii) 740 nm.}
\label{Fig2}
\end{figure}

\begin{figure*}[t]
\centerline{\includegraphics[width=17.0cm]{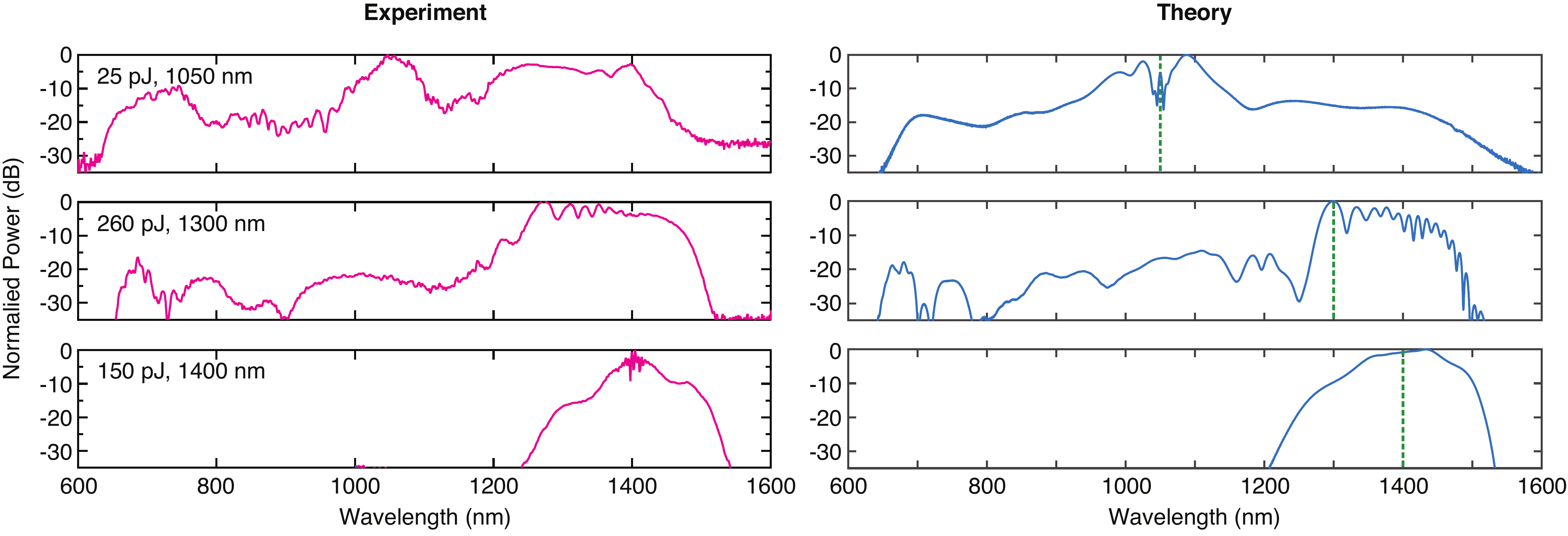}}
\caption{Experimental (left) and the corresponding theoretical (right) spectra for supercontinuum generation in a 730$\times$700 nm Si$_3$N$_4$ waveguide for three different pump regimes are considered. A pump wavelength of 1050 nm (top) corresponds to anomalous GVD, 1300 nm corresponds to normal GVD, and 1400 nm corresponds to large normal GVD.}
\label{Fig3}
\end{figure*}

We theoretically consider the pulse propagation dynamics in the normal GVD regime (Fig. \ref{Fig1}). We simulate the dynamics via a split-step Fourier method to solve the nonlinear Schr\"{o}dinger equation, which includes effects of third-order nonlinearity, higher-order dispersion (HOD), and self-steepening \cite{Johnson}. We assume 200-fs pulses centered at 1300 nm with 1.3 kW of peak power (260 pJ pulse energy). We use a 2-cm-long Si$_3$N$_4$ waveguide with a cross section of 730$\times$700 nm, that yields the GVD profile as shown in Fig. \ref{Fig2}(a) which is characterized by zero-GVD wavelengths at 890 nm and 1178 nm and exhibits normal GVD at 1300 nm. We consider the dynamics under two different conditions. In order to verify that DW generation is due to HOD, we first consider only the contribution from the GVD and neglect HOD terms. Figure \ref{Fig1}(a) and (b) shows the simulated spectral evolution and the spectrogram at the waveguide output, respectively. Here, we observe spectral broadening due to self-phase modulation (SPM) along with temporal pulse broadening as seen in the spectrogram. While the spectrum spans from 1000 nm to 1600 nm, in the absence of HOD, the SCG spectrum is largely symmetric and dispersive wave generation does not occur. The spectral evolution and spectrogram with the inclusion of HOD is shown in Fig. \ref{Fig1}(c) and (d), respectively. Similar to the case without HOD, we see that the spectrum initially broadens due to SPM. However, at longer propagation lengths ($z=14$ mm), we observe depletion of the low wavelength component of the pump below 1255 nm and significant spectral components generated towards shorter wavelengths (650 nm). In addition, the spectrogram indicates that while the spectral components near the pump wavelength undergo temporal broadening, pulse compression occurs near 1 $\mu$m. Figure \ref{Fig2}(b) shows the simulated spectrum at the output. We clearly see significant depletion of the original pump pulse spectrum below 1300 nm. Furthermore, we observe the formation of two peaks near 685 nm and 740 nm, which we attribute to DW formation, and the generated spectrum spans over an octave of bandwidth.         

The dynamics of DW generation can be described as a phase-matched cascaded four-wave mixing process \cite{Foster,Erkintalo}, such that the phase matching condition governed by the dispersion operator \cite{Dudley},

\begin{equation}
\hat{D} = \sum_{\mathclap{n=2,3,...}}\frac{\beta_n(\omega_0)}{n!}(\omega-\omega_0)^n = 0,
\end{equation}

\noindent
where $\beta_n$ is the $n$\textsuperscript{th}-order dispersion coefficient, and $\omega_0$ is the pump frequency. Here, we assume the nonlinear contribution to the phase mismatch is negligible \cite{Webb}. The dispersion operator \emph{\^{D}} is plotted in Fig. \ref{Fig2}(c) for two different pump wavelengths. For a pump wavelength of 1255 nm (magenta) [(iv) in Fig. \ref{Fig2}(c)], the spectral component at 1255 nm is converted to 685 nm and 1 $\mu$m [(i) and (iii) in Fig. \ref{Fig2}(c), respectively]. The 685-nm component corresponds to the short wavelength DW, and the 1-$\mu$m component lies in the anomalous-GVD regime. For a pump wavelength of 1 $\mu$m, the dispersion operator (cyan) yields phased-matched generation of a component at 740 nm, which agrees with the simulated spectrum. Thus, the generation of this peak is due to a cascaded DW process that is enabled by pulse compression of the first dispersive wave near 1 $\mu$m. 

\begin{figure}[tbp]
\centerline{\includegraphics[width=8.5cm]{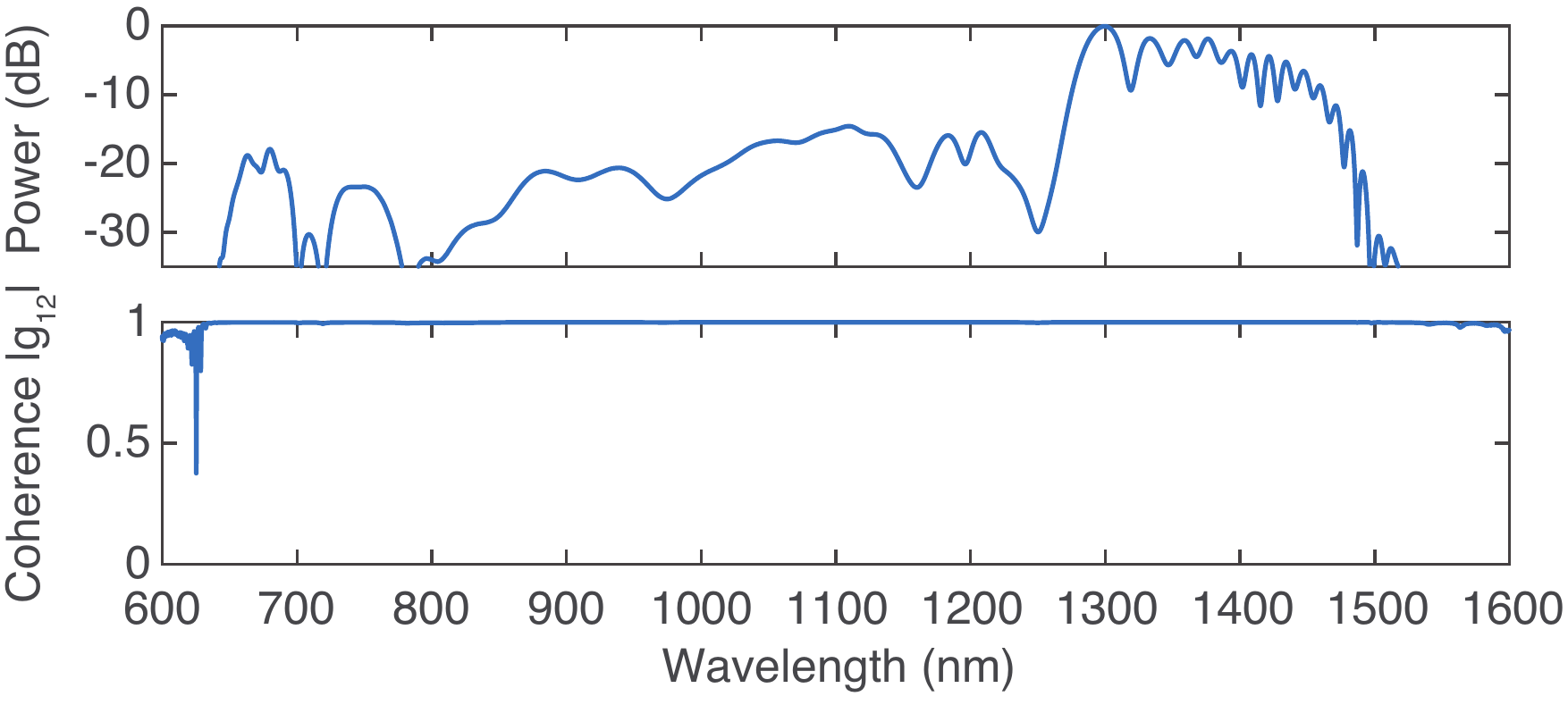}}
\caption{Simulated spectrum (top) and calculated first-order mutual coherence (bottom) for SCG pumped in normal-GVD regime for an input pulse at 1300 nm.}
\label{Fig4}
\end{figure}

We investigate pumping in different dispersion regimes, both experimentally and theoretically (Fig. \ref{Fig3}). In our experiment, the pump source is a 200-fs pulse train from a tunable optical parametric oscillator (OPO) with a repetition rate of 80 MHz. We pump a 2-cm-long Si$_3$N$_4$ waveguide with a 730$\times$700 nm cross section with OPO pulses at 1050 nm, 1300 nm, or 1400 nm, which corresponds to anomalous GVD ($\beta_2 = -0.036$ ps$^2$/m), normal GVD ($\beta_2 = 0.108$ ps$^2$/m), and larger normal GVD ($\beta_2 = 0.27$ ps$^2$/m), respectively. The pulse energies in the waveguide are calculated based on the measured output power based on a fixed incident average power of 120 mW, and are 25 pJ, 260 pJ, and 150 pJ for pump wavelengths of 1050 nm, 1300 nm, and 1400 nm, respectively. The variations are due to variations in the coupling efficiencies for different wavelengths. For a 1050-nm pump (anomalous GVD), we observe a pair of DW's form in the normal-GVD regime at 700 nm and 1300 nm [Fig. \ref{Fig3}(a)]. For the 1300-nm pump (normal GVD), we observe a 1.2-octave-spanning SCG spectrum with two DW peaks at 685 nm and at 770 nm. In contrast, for a 1400-nm pump (large normal GVD), we only observe spectral broadening due to SPM. While the spectral coverage pumping at 1050 nm and 1300 nm is similar, pumping in the anomalous-GVD regime (1050 nm) results in the degradation of the coherence due to modulation instability \cite{Nakazawa,Demircan}, and limits the allowable pump power. In contrast, pumping in normal-GVD regime (1300 nm) relies on optical wave breaking for broadening and allows for low noise and high coherence with higher pump powers \cite{Millot}. Additionally, we simulate the output SC spectra pumping with conditions similar to our experiment. The simulations shows good agreement with the measured SCG spectra. At 1400 nm, we observe only SPM broadening at pulse energies of 150 pJ and up to a higher pulse energy of 260 pJ. In contrast, at 1300 nm and 260 pJ, cascaded DW formation occurs. We believe the difference in the spectra between theory and experiment are largely a result of deviations between the actual dispersion and simulated dispersion due to waveguide fabrication tolerances. Our results confirm that, with dispersion engineering, we can take advantage of the two zero-GVD points to produce broadband coherent SCG by pumping in the normal-GVD regime, which results in a highly asymmetric spectrum that is directional to the blue with respect to the pump frequency. 

\begin{figure}[tbp]
\centerline{\includegraphics[width=7.0cm]{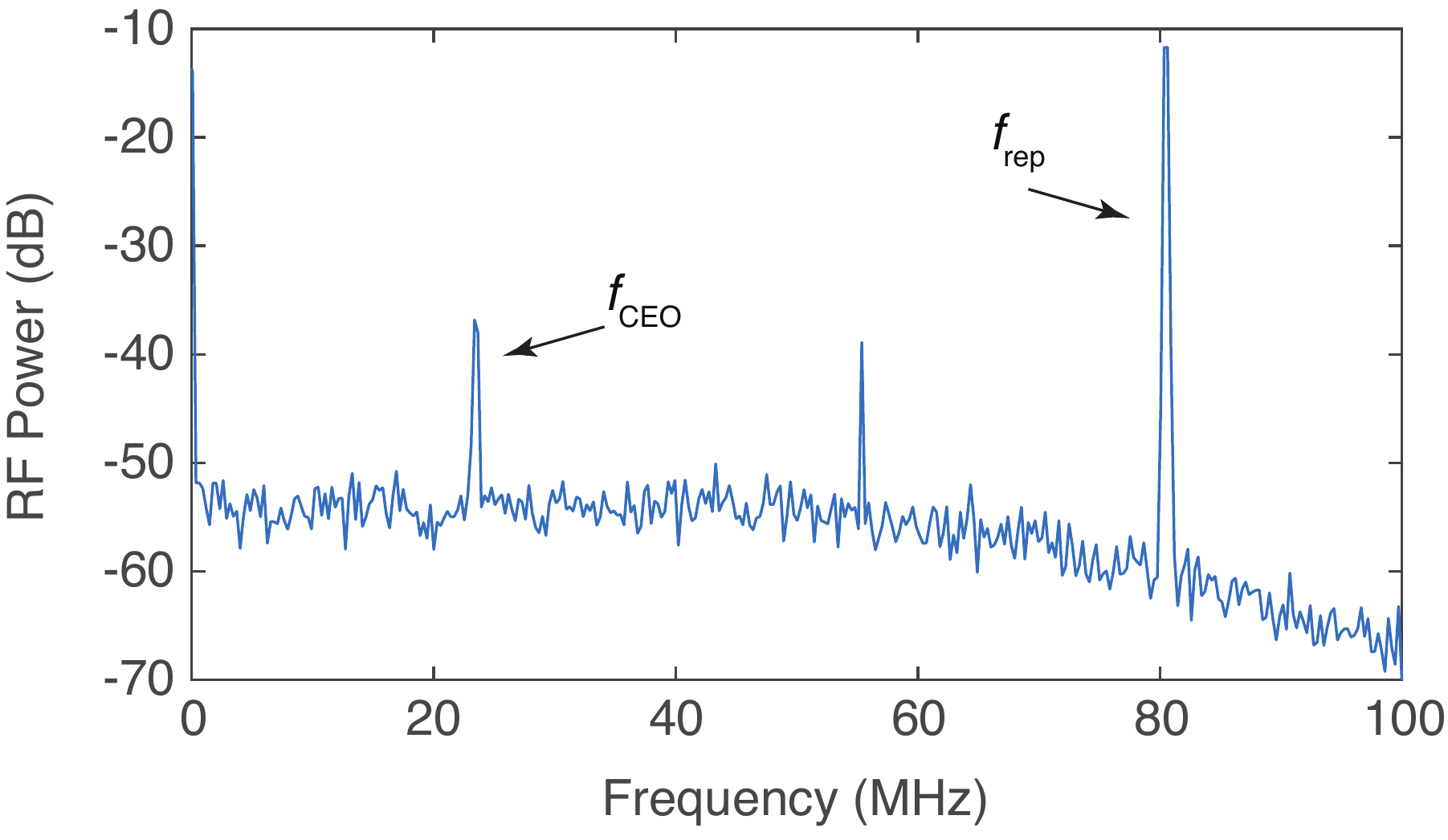}}
\caption{Measured $f\textsubscript{CEO}$ using a \emph{f}-2\emph{f} interferometer at 690 nm. The $f\textsubscript{CEO}$ is measured at 23 MHz with a signal-to-noise ratio of 17 dB.} 
\label{Fig5}
\end{figure}

We investigate the spectral coherence numerically for a 1300-nm pump in which the equivalent to quantum shot noise is added to our input pulse \cite{Ruehl}, and the first-order mutual coherence function $g_{12}$\cite{Johnson,Gu} is calculated. Figure \ref{Fig4} shows the simulated spectrum along with the calculated spectral coherence, and the spectrum exhibits a high degree of coherence over the entire bandwidth. 

We demonstrate the coherence properties of the SCG by performing $f\textsubscript{CEO}$ detection using an \emph{f}-2\emph{f} interferometer similar to \cite{Mayer}. The waveguide output is sent to a Michelson interferometer with a dichroic beamsplitter, which allows for tuning the time delay between the short and long wavelength components of the generated SC spectrum. The output of the interferometer is sent to a 4-cm-long periodically-poled lithium niobate (PPLN) crystal to allow for frequency doubling of the 1380 nm component. The PPLN output is spectrally filtered using a bandpass filter centered at 690 nm with a 10-nm bandwidth and sent to a photodiode and an RF spectrum analyzer (see Fig. \ref{Fig5}). The repetition frequency $f\textsubscript{rep}$ is 80 MHz, and we observe the $f\textsubscript{CEO}$ at 23 MHz with a signal-to-noise ratio of 17 dB. The signal-to-noise of the $f\textsubscript{CEO}$ beat can be further improved by optimizing spatial collimation of the waveguide output for the \emph{f}-2\emph{f} wavelengths. This measurement clearly shows the high coherence properties of the spectra generated via this cascaded DW process.

In conclusion, we demonstrate coherent SCG in a Si$_3$N$_4$ waveguide pumped in the normal-GVD regime through cascaded DW formation. Through suitable dispersion engineering of the waveguide, we generated a highly asymmetric SC spectrum towards the blue that spans over an octave of bandwidth. We verify the coherence of the generated spectrum numerically and through $f\textsubscript{CEO}$ detection using an \emph{f}-2\emph{f} interferometer. Our technique offers a path towards a high-power coherent SC source with applications including optical clocks, precision metrology, spectroscopy, and optical frequency synthesis. We expect that this can be applied to other wavelength regimes, including the generation of spectra towards the mid-infrared.

\textbf{Funding.} Defense Advanced Research Projects Agency (DARPA) (N66001-16-1-4055); Air Force Office of
Scientific Research (AFOSR) (FA9550-15-1-0303); National Science Foundation (NSF) (ECS-0335765).
\\
\\
\textbf{Acknowledgment.} This work was performed in part at the Cornell Nano-Scale Facility, which is a member of the National Nanotechnology Infrastructure Network, supported by the NSF, and at the CUNY Advanced Science Research Center NanoFabrication Facility. We also acknowledge useful discussions with A. Klenner. 


\clearpage
\newpage

\end{document}